
\input harvmac
\noblackbox
\def\IR{{\hbox{{\rm I}\kern-.2em\hbox{\rm R}}}}
\def\IB{{\hbox{{\rm I}\kern-.2em\hbox{\rm B}}}}
\def\IN{{\hbox{{\rm I}\kern-.2em\hbox{\rm N}}}}
\def\IC{{\ \hbox{{\rm I}\kern-.6em\hbox{\bf C}}}}

\font\cmss=cmss10 \font\cmsss=cmss10 at 7pt
\def\IZ{\relax\ifmmode\mathchoice
{\hbox{\cmss Z\kern-.4em Z}}{\hbox{\cmss Z\kern-.4em Z}}
{\lower.9pt\hbox{\cmsss Z\kern-.4em Z}}
{\lower1.2pt\hbox{\cmsss Z\kern-.4em Z}}\else{\cmss Z\kern-.4em Z}\fi}

\def\NP{{\it Nucl. Phys.\ }}

\def\PL{{\it Phys. Lett.\ }}
\def\PR{{\it Phys. Rev.\ }}
\def\PRL{{\it Phys. Rev. Lett.\ }}

\def\Mod{{\it Mod. Phys. Lett.\ }}

\def\muh{\hat \mu}
\def\nuh{\hat \nu}
\def\phih{\hat \phi}

\baselineskip 12pt
\Title{\vbox{\baselineskip12pt\hbox{hep-th/9403186}\hbox{UCSBTH-94-14}}}
{\vbox{\hbox{\centerline{Exact Four-Dimensional
Dyonic Black Holes}}
\hbox{\centerline{and Bertotti-Robinson Spacetimes in String
Theory}}}}
\centerline{David A. Lowe\footnote*{lowe@tpau.physics.ucsb.edu}}
\centerline{and}
\centerline{Andrew Strominger\footnote\dag{andy@denali.physics.ucsb.edu}}
\smallskip
\centerline{Department of Physics}
\centerline{University of California}
\centerline{Santa Barbara, CA 93106-9530}
\smallskip
\vskip .5cm
\noindent
Conformal field theories corresponding to  two--dimensional
electrically charged black holes and to  two--dimensional anti--de Sitter
space with a covariantly constant electric field are simply constructed
as  $SL(2,\IR)/ \IZ$  WZW coset models. The two--dimensional
electrically charged black holes are related by
Kaluza--Klein reduction to the 2+1--dimensional rotating black hole
of Ba\~nados, Teitelboim and Zanelli, and our construction is
correspondingly related to its realization as a WZW model.
Four--dimensional
spacetime solutions are obtained by tensoring these  two--dimensional theories
with $SU(2)/Z(m)$ coset models. These describe
a family of dyonic black holes and the Bertotti--Robinson
universe.

\Date{March, 1994}

\newsec{Introduction}

\lref\witten{E. Witten,\PR
{\bf D44} (1991) 314.}

\lref\mbh{G.W. Gibbons, \NP {\bf B207} (1982) 337;
G.W. Gibbons and K. Maeda, \NP {\bf B298}
(1988) 741;
 D. Garfinkle, G. Horowitz, and A. Strominger, \PR {\bf D43} (1991) 3140.}%

\lref\gps{S. Giddings, J. Polchinski and A. Strominger, \PR {\bf D48}
(1993) 5784.}%

\lref\will{W. Nelson, University of California at Santa Barbara
preprint,UCSBTH-93-10.}%

\lref\asup{A. Strominger, unpublished.}

\lref\kallosh{R. Kallosh, A. Linde, T. Ortin, A. Peet and
A. Van Proeyen, \PR {\bf D46} (1992) 5278;
R. Kallosh and A. Peet, \PR {\bf D46} (1992) 5223.}%

\lref\nappi{C. Nappi, M.D. McGuigan and S. Yost, \NP {\bf B375} (1992) 421.}

\lref\achu{A. Ach\'ucarro and M. Ortiz, \PR {\bf D48} (1993) 3600.}

\lref\banad{M. Ba\~nados, C. Teitelboim, J. Zanelli,
\PRL {\bf 69} (1992) 1849 ;  M. Ba\~nados, M. Henneaux, C. Teitelboim, J.
Zanelli,
\PR {\bf D 48} (1993) 1506.}

\lref\gist{S. Giddings and A. Strominger, \PR {\bf D46} (1992) 627.}

\lref\horo{G. Horowitz and D. Welch, \PRL
{\bf 71} (1993) 328; N. Kaloper, \PR {\bf D48} (1993) 2598.}

\lref\wilc{A. Shapere, S. Trivedi and F. Wilczek, \Mod {\bf A6} (1991) 337.}

Near extremality, four--dimensional, magnetically charged
black holes in string theory
develop a large cigar--shaped region adjacent to the horizon
\mbh\ . The solution in
this region degenerates into a product of two two--dimensional factors,
both of which can be described as exact conformal field theories and hence
as exact solutions of string theory.
One of these factors is Witten's two-dimensional black hole \witten , and the
other
is a two--sphere with magnetic charge and is represented by a WZW
orbifold \refs{\gps, \will}. At extremality, the two--dimensional black hole
factor
is replaced by the linear dilaton vacuum.

It is known from analysis of the low--energy effective field theory that
the addition of electric charge dramatically changes the
picture \refs{\mbh,\wilc ,\kallosh}.
The solutions still degenerate into two factors near the horizon, but the
two--dimensional black hole factor of \witten\ must be replaced by an
asymptotically anti--de Sitter, electrically charged black hole
similar to those
discussed in \nappi . In the extremal limit, the black hole factor becomes
electrically charged anti-de Sitter space. Including the magnetically charged
two--sphere, one then has the Bertotti--Robinson universe. This limiting
behavior is identical to that of Reissner-Nordstr{\o}m black holes.

In this paper, we give an exact construction of the electrically charged
two--dimensional black hole, and corresponding family of
four--dimensional dyonic black holes. This is rather easily
accomplished by
using the observation of Ach\'ucarro and Ortiz \achu\
that the  2+1--dimensional rotating black hole
of Ba\~nados, Teitelboim and Zanelli \banad\ may be
reinterpreted as an electrically charged
two--dimensional black hole. We may then apply the exact construction of
those black holes by Horowitz, Welch and Kaloper \horo\ to our case.
Related constructions give two--dimensional electrically
charged anti-de Sitter space and  the Bertotti--Robinson universe.

\newsec{Kaluza--Klein reduction from 3 to 2 dimensions}
\subsec{Two--dimensional charged black holes}

Ba\~nados et al.
\banad\
have constructed a family of spinning black holes
in $2+1$--dimensional Einstein gravity with negative
cosmological constant proportional to $l^2$.
For general values of the mass $M$ and angular momentum $J$,
the $2+1$--dimensional spacetime metric is
\eqn\tpomet{
ds^2 = \biggl( M - {{ r^2} \over {l^2}} \biggr) dt^2
- J dt d \varphi + r^2 d \varphi^2 + \biggl( {{ r^2} \over {l^2}}
-M + { {J^2 } \over {4 r^2}} \biggr)^{-1} dr^2 ~,
}
where $\varphi$ is periodically identified with period
$2\pi$, and $r$ and $t$ may take any real value. There
are two horizons located at $r=r_{\pm}$, where
$r_{\pm}$ are related to $M$ and $J$ via
\eqn\inout{
M= { {r_+^2 + r_-^2} \over {l^2}}, \qquad J={{2 r_+ r_-} \over l}~.
}
The above identification breaks down in the extreme case $|J|= M l$,
when the two horizons coincide, and also in the massless case
$M=J=0$. In these situations, the periodic identification must be made
along a null direction.

The 2+1 black hole \tpomet\ also corresponds to a  solution of string theory
with a non--trivial antisymmetric tensor \horo .
The relevant part of the three--dimensional string effective action is
\eqn\threffac{
S_3 = \int d^3 x \sqrt{-\hat g} e^{-2 \hat \phi} \biggl[
{4 \over l^2} +\hat R +
4 (\nabla \hat \phi)^2
- {1 \over 12} {{\hat H}^2}
\biggr] ~,}
where $3l^2/(l^2-2)$ is the central charge
of the conformal field theory.
The antisymmetric tensor field strength for
the black hole solutions is
\eqn\asten{
\hat H_{\muh \nuh \hat \lambda} = {2 \over l}~
\epsilon_{\muh \nuh \hat \lambda}~,
}
where $\muh$ runs over $r,t, \varphi$.

Here we would like to reinterpret this three--dimensional solution as a
two--dimensional charged black hole. This is achieved,
following Ach\'ucarro and Ortiz,
by performing a Kaluza--Klein reduction of the metric
\tpomet\ and antisymmetric tensor,
regarding $\varphi$ as the compact coordinate
which will generate a $U(1)$ gauge field. The
$2+1$--dimensional spacetime metric is written
in terms of the two--dimensional metric $g_{\mu \nu}$, $U(1)$ gauge
field $A^{}_{\mu}$,
and scalar $D$
\eqn\tpokkmet{
\hat g = \left( \matrix{ g_{\mu \nu}+ e^{2D} A^{}_{\mu}
A^{}_{\nu } & e^{2D} A^{}_{\mu } \cr
e^{2D} A^{}_{\nu} & e^{2D} \cr} \right)~.
}
Here
$\mu,\nu$ run over $r,t$ and unhatted quantities refer to
two--dimensional fields.
The two--dimensional low--energy effective action is
\eqn\effac{
S_2 = \int d^2 x \sqrt{-g} e^{D-2 \phih} \biggl[ {4 \over l^2} +R +
4 (\nabla \phih)^2
-4 { {\nabla D \cdot \nabla \phih } }- {1 \over 4}e^{2D} F^2- {1 \over 4}
e^{-2D}F^{\prime 2}
\biggr] ~,
}
where $F$ ($F^{\prime}$) is the field strength constructed from
$A_\mu$ ($\hat B_{\mu \varphi}$).
The antisymmetric tensor $B_{\mu \nu}$ drops out of this expression
and will henceforth be ignored.

It is possible to eliminate one
of the gauge fields and the dilaton in a manner consistent with the
equations of motion. This is achieved by setting
\eqn\gtrunc{
\phih =  { 0,} \qquad
F^{\prime}_{\mu \nu} = { {2 e^D} \over l} \epsilon_{\mu \nu}~,
}
where $\epsilon_{\mu \nu}$ is the two--dimensional volume form.
The truncated equations of motion then follow from
\eqn\effactr{
S_2 = \int d^2 x \sqrt{-g} e^{D} \biggl[ {2\over l^2} +R
 - {1 \over 4}
e^{2D} F^{ 2}
\biggr] ~.}
The two--dimensional spacetime
solutions we consider are all compatible with this truncation.
\effactr\ is the Jackiw--Teitelboim model with a gauge field
\ref\rjack{R. Jackiw, in {\it Quantum Theory of Gravity}, ed.
S. Christensen, Hilger, Bristolm 1984;
C. Teitelboim, {\it ibid}.}.

The Kaluza--Klein reduction of the three--dimensional
metric \tpomet\ and antisymmetric
tensor field leads to
\eqn\tdmet{
ds^2 = \biggl( M - {{ r^2} \over {l^2}} - { {J^2 } \over {4 r^2}}
\biggr) dt^2
+ \biggl( {{ r^2} \over {l^2}}
-M + { {J^2 } \over {4 r^2}} \biggr)^{-1} dr^2~,
}
with $U(1)$ gauge field and scalar
\eqn\gfield{
A^{}_t = - {J \over {2 r^2}}~, \qquad D= \log r ~,
}
which is of course an extrema of \effactr.\foot{Note that
decompactification ($D \rightarrow \infty$) occurs
at large $r$. However, in order to obtain an asymptotically
flat four--dimensional black hole one must tensor with a two--sphere
as discussed in subsequent sections and
``sew on'' an asymptotically
flat four--dimensional region at some finite radius from the
horizon.
The growth of $D$ will then be cut off at the
sewing radius.
This sewing procedure is readily understood in the
low--energy effective field theory \refs{\kallosh,\gist}, but not in
the conformal field theory, where it presumably
corresponds to perturbation by a ``bad'' marginal operator.}

Evaluating the Ricci
scalar for the metric \tdmet\ gives
\eqn\ricci{
R= - {{4 r^4 + 3 J^2 l^2} \over {2 l^2 r^4}}~.
}
The two--dimensional black hole metric is asymptotically anti--de Sitter, with
a curvature singularity at $r=0$.

Horowitz, Welch and Kaloper
\horo\ have shown that the solution \asten , \tpomet ,
corresponds to the $SL(2, \IR)$ WZW model with
a discrete identification and is therefore exact. The $SL(2,\IR)$ WZW action is
invariant under
the following symmetries
\eqn\wzwsym{
\delta g = \epsilon \left[ \left( \matrix{ 1&0 \cr 0 &-1 \cr} \right)g
+ g \left( \matrix{1&0 \cr 0 & -1} \right) \right]
}
and
\eqn\wzwsy{
\delta g = \epsilon \left[ \left( \matrix{ 1&0 \cr 0 &-1 \cr} \right)g
- g \left( \matrix{1&0 \cr 0 & -1} \right) \right]~,
}
where $g$ is an element of $SL(2,\IR)$.
Quotienting $SL(2,\IR)$ by a discrete subgroup of a linear combination
of these symmetries yields the general two parameter family of
black holes \tpomet. The level $k_{SL}$ of the WZW model is fixed by
\eqn\lev{
k_{SL} = l^2~.
}

\subsec{1+1--dimensional anti--de Sitter space with constant electric field}

Anti--de Sitter space in 1+1 dimensions with a covariantly constant
electric field (not considered in \horo ) may be obtained
by a similar construction which we will describe in some detail. The difference
here will be that the periodic identification of 2+1--dimensional
anti--de Sitter space ({\it ie} $SL(2,\IR)$) will be made
along a direction that is always spacelike. In the previous
section the identification was made along a Killing vector
that was timelike in certain regions.

It is convenient to
use  coordinates which differ from those of the previous subsection
to describe the
anti--de Sitter solution.
Here we parametrize an element of $SL(2,\IR)$, with all elements non--zero,
by \ref\vilen{N. Vilenkin and A. Klimyk, ``Representation of Lie Groups
and Special Functions,'' Kluwer Academic Publishers (1991).}
\eqn\sltwo{
g = d_1 (-e)^{\epsilon_1} s^{\epsilon_2} p d_2~,
}
where $\epsilon_1, \epsilon_2 = 0$ or 1,
\eqn\stuff{
\eqalign{
d_1 &= \left(\matrix{ e^{t_L/2} & 0 \cr
                      0 &e^{-t_L/2} \cr}\right)~, \qquad
d_2 = \left(\matrix{ e^{(-1)^{\epsilon_2}{t_R/2}}&0\cr
                         0& e^{- (-1)^{\epsilon_2}t_R/2}\cr}\right)~, \cr
-e &= \left( \matrix{ -1 & 0 \cr
                      0&-1\cr}\right)\qquad
s = \left( \matrix{ 0 & 1 \cr
                    -1 & 0 \cr}\right)~, \cr}
}
and $p$ is either
\eqn\sstuff{
\eqalign{
(i)\qquad p &= \left( \matrix{ \cosh(r/2) & \sinh(r/2) \cr
                    \sinh(r/2)& \cosh(r/2) \cr }\right)~,\quad
-\infty < r < \infty \quad {\rm or} \cr
(ii) \qquad p&= \left( \matrix{ \cos(r/2) & \sin(r/2) \cr
                    -\sin(r/2)& \cos(r/2) \cr }\right)~,
 \quad -\pi/2<r<\pi/2~,\cr}
}
with $t_L , t_R \in \IR$.

The WZW action  is
\eqn\wznwact{
S_{WZW} = {{ k_{SL}} \over {4\pi}} \int_{\Sigma} d^2 z {\rm Tr} (g^{-1}
\partial
g g^{-1} \bar \partial g ) +  k_{SL} \Gamma(g) ~,
}
where $\Gamma$ is the Wess Zumino term
\eqn\wessz{
\Gamma = {1 \over {12 \pi}} \int_B {\rm Tr} ( g^{-1} d g \wedge
g^{-1} d g \wedge g^{-1} d g )~.
}
Here $\Sigma$ is a Riemann surface and $B$ is a 3--manifold
with boundary $\Sigma$.
With $p$ in region (i), \wznwact\
takes the form
\eqn\wzwparam{
S_{WZW} = {{ k_{SL}} \over {4\pi}} \int  ( \partial r \bar \partial r
+ \partial t_L \bar \partial t_L + \partial t_R \bar \partial t_R
+  2 \cosh r ~\bar \partial t_L  \partial t_R ) ~.
}
With $p$ in region (ii) the answer is
\eqn\wzwparamtwo{
S_{WZW} = {{ k_{SL}} \over {4\pi}} \int  ( -\partial r \bar \partial r
+ \partial t_L \bar \partial t_L + \partial t_R \bar \partial t_R
+  2 \cos r ~\bar \partial t_L  \partial t_R )~.
}

This may be viewed as a $\sigma$--model on a three--dimensional
target space with metric $\hat g_{\muh \nuh}$ and
antisymmetric tensor $\hat B_{\muh \nuh}$. One finds
in region (i)
\eqn\kkmetric{
\hat g = {1\over 4}\left( \matrix{ k_{SL}&0&0 \cr
0& k_{SL} & k_{SL} \cosh r \cr
0& k_{SL} \cosh r & k_{SL} \cr
} \right)~,
}
with the ordering of coordinates taken to be
$(r, t_L,  t_R)$.
The non-zero component of the antisymmetric tensor is
\eqn\kkas{
\hat B_{t_L t_R} = {1\over 4} k_{SL} \cosh r~. }
In region (ii) one finds
\eqn\tkkmetric{
\hat g = {1\over 4} \left( \matrix{ -k_{SL}&0&0\cr
0& k_{SL} & k_{SL} \cos r \cr
0& k_{SL} \cos r & k_{SL} \cr
} \right)~,
}
with antisymmetric tensor
\eqn\tkkas{
\hat B_{t_L t_R} = {1\over 4} k_{SL} \cos r~. }

The WZW action is invariant under
translations in $t_R$ which are generated by
\eqn\trtrans{
\delta g = \epsilon g \left( \matrix{ 1 & 0 \cr
                                      0 & -1 \cr}\right)~.
}
We  quotient
$SL(2,\IR)$ by a discrete subgroup of this
symmetry: the integers
$\IZ$. $t_R$ becomes a periodic coordinate with
range $[0, 2\pi/\gamma)$, where $\gamma$ is a real parameter.

We may now perform a Kaluza--Klein reduction of the metric
and antisymmetric tensor, to obtain a two--dimensional
spacetime with a $U(1)$ gauge field in the same way as
before. The identifications of \tpokkmet\
give the fields in region (i)
\eqn\gfields{
\eqalign{
A^{}_{t_L } &=  \gamma  \cosh r~, \qquad
e^{2D} = {{k_{SL}}\over {4 \gamma^2}}~, \cr
g &= {1\over 4} \left( \matrix{ k_{SL} &0 \cr
0& - k_{SL} \sinh^2 r  \cr
} \right)~. \cr}
}

Similarly
in region (ii) the fields are
\eqn\gtfields{
\eqalign{
A^{}_{t_L } &= \gamma  \cos r~, \qquad
e^{2D} = {{k_{SL}}\over {4 \gamma^2}}~, \cr
g &= {1\over 4} \left( \matrix{ -k_{SL} &0 \cr
0&  k_{SL} \sin^2 r  \cr
} \right)~. \cr}
}

We now show this corresponds to a double cover of
two--dimensional anti--de Sitter space, $(adS)_2$.
Recall $(adS)_2$ may be defined
as the hyperboloid
\eqn\adsdef{
-x_0^2 - x_1^2 + x_2^2 = -1~,
}
embedded in flat three--dimensional space with metric
\eqn\adsmet{
ds^2 = - dx_0^2 - dx_1^2 + dx_2^2~.
}
Now define coordinates
\eqn\ricoord{
\eqalign{
x_0 &= \cosh r \cr
x_1 &= \sinh r \sinh t_L \cr
x_2 &= \sinh r \cosh t_L ~,\cr}
}
which covers the region $x_0>1$ when $r,t_L \in  \IR$. The metric in these
coordinates is
\eqn\rimet{
ds^2 = dr^2 - \sinh^2 r dt_L^2~,
}
as in region (i), up to a rescaling by $k_{SL}/4$.
Remember we actually have 4 copies
of region (i) corresponding to $\epsilon_1 = 0, 1$ and
$\epsilon_2 = 0, 1$. Two of these copies are sufficient
to cover the region $|x_0| > 1$.

To cover the region $0<x_0< 1$ define coordinates
\eqn\riicoord{
\eqalign{
x_0 &= \cos r \cr
x_1 &= \sin r \cosh t_L \cr
x_2 &= \sin r \sinh t_L ~,\cr}
}
where $t_L \in \IR$ and $r \in (-\pi/2 , \pi/2)$.
The metric is then
\eqn\riimet{
ds^2 = -dr^2 + \sin^2 r dt_L^2~,
}
agreeing with the metric above in region (ii). A second
copy of region (ii) will cover $-1<x_0<0$.
We find then that the WZW model corresponds to
a double cover of $(adS)_2$.
It may also
be verified that the gauge field strength is covariantly
constant with respect to this metric.

\newsec{Four--dimensional exact solutions}

Having obtained the above two--dimensional spacetime solutions
it is a simple matter to tensor these with the angular
magnetic monopole CFT of \refs{\gps, \will} to obtain four--dimensional
spacetime solutions with non--trivial electric and
magnetic fields.
The monopole CFT is obtained by quotienting a
$SU(2)$ WZW model by the discrete subgroup $Z(m)$.

To parametrize the $SU(2)$ group manifold
we introduce the real coordinates
$\theta, \phi$ and $\zeta$ and write the group element $g'$ as
\eqn\sutwo{
g' = e^{{i\over 2} \phi \sigma_3} e^{ {i\over 2} \theta \sigma_1 }
e^{ {i\over 2} \zeta \sigma_3}~,
}
where $\theta \in [0,\pi]$, $\phi \in [0,2\pi]$ and $\zeta \in [0,4\pi]$,
and the $\sigma_i$ are the Pauli matrices.
The $SU(2)$ WZW action is then
\eqn\wzwsu{
S_{WZW} =
 {{ k_{SU} } \over {4\pi}} \int  ( \partial \theta \bar \partial \theta
+ \partial \phi \bar \partial \phi+ \partial \zeta \bar \partial \zeta
+ 2 \cos \theta  ~\bar \partial \phi \partial \zeta) ~.
}

As in \will, this three--dimensional sigma model action may be
compactified to 2 dimensions. The extra coordinate $\zeta$, leads
to a $U(1)$ gauge field corresponding to a magnetic field, while
the two--dimensional space is just $S^2$.
Quotienting the $SU(2)$ manifold by a discrete subgroup $Z(m)$,
identifying $\zeta \sim \zeta + {{4\pi} \over m}$,
determines the charge of the magnetic monopole $Q=m$.
Modular invariance
requires $k_{SU} = n m$ where $n$ and $m$ are integers.
(Alternatively one could use the conformal field theories
of \gps, which involve magnetic charge associated with a general
$U(1)$ current algebra.)

\subsec{Four--dimensional charged black holes}

Tensoring this monopole CFT with the CFT described in
section 2.1 leads to a solution describing the throat limit
of a four--dimensional black hole with electric and magnetic charge.
The four--dimensional spacetime is identified with the
$r,t, \theta$ and $\phi$ coordinates. The two compactified
dimensions
correspond to the $\zeta$ and $\varphi$ coordinates.

In the following $A^{\varphi}_{\mu}$ ($A^{\zeta}_{\mu}$) refers
to the gauge field derived from the $\hat g_{\mu \varphi}$
($\hat g_{\mu \zeta}$) components of the six--dimensional metric.
Also, $A'_{\mu, \varphi}$ ($A'_{\mu, \zeta}$) refers
to the gauge field constructed from the $\hat B_{\mu \varphi}$
($\hat B_{\mu \zeta}$) components of the six--dimensional
antisymmetric tensor. The scalar field $D_{\varphi}$ ($D_{\zeta}$)
is derived from the $\hat g_{\varphi \varphi}$ ($\hat g_{\zeta \zeta}$)
component of the metric.

Note that in four--dimensions we may no longer truncate
the gauge field $A'_{\mu, \varphi}$ and dilaton as in \gtrunc.
However the gauge field $A'_{\mu, \zeta}$ and the scalar $D_{\zeta}$
may be truncated with the identification
\eqn\btrunc{
D_{\zeta} = {\rm const,} \qquad
A'_{\mu,\zeta} = \pm  e^{2D_{\zeta}} A^{\zeta}_{\mu}~,
}
in a manner consistent with the equations of motion.
The non--trivial components of the spacetime fields are then
\eqn\cbhfour{
\eqalign{
ds^2 &= \biggl( M - {{ r^2} \over {l^2}} - { {J^2 } \over {4 r^2}}
\biggr) dt^2
+ \biggl( {{ r^2} \over {l^2}}
-M + { {J^2 } \over {4 r^2}} \biggr)^{-1} dr^2
+ {1\over 4} k_{SU} d \theta^2 +
{1\over 4}k_{SU} \sin^2 \theta d \phi^2~, \cr
A^{\varphi}_t &= - {J \over {2 r^2}},\qquad
A'_{t, \varphi} = - {{r^2 }\over l}~,
\qquad A^{\zeta}_{\phi} =
m \cos \theta,  \qquad
\phih = {\rm const}, \cr
e^{2D_{\varphi}} &= r^2, \qquad e^{2D_{\zeta}} = {n \over {4m}},
\qquad H_{\mu \nu \rho} =0 ~.\cr}
}

Since we wish to identify this with a six--dimensional
Kaluza--Klein spacetime, the central charge should be 6, i.e.
\eqn\ccharge{
c= {{3k_{SU}} \over {k_{SU}+2}} + {{3 k_{SL}} \over {k_{SL}-2} } =6~.
}
This leads to the condition $k_{SL}= k_{SU}+4$. Taking
$k_{SL}$ and $k_{SU}$ both positive leads to a spacetime
with the correct $(-+++)$ signature.

\subsec{Bertotti--Robinson Universe}

If instead we tensor the monopole CFT with the anti--de Sitter
solution of section 2.2, we will obtain, in the large $k_{SU}$ and
large $k_{SL}$ limit, the four--dimensional
Bertotti--Robinson spacetime \ref\bert{B. Bertotti, \PR {\bf 116} (1959) 1331;
I. Robinson, {\it Bull. Acad. Polon. Sci.} {\bf 7} (1959) 351.}
with covariantly constant electric and magnetic fields.
This solution describes the throat limit of
the extremal dilaton black holes with electric and
magnetic charge described in
\kallosh,
and also the throat limit of the familiar extremal
Reissner--Nordstr{\o}m black hole.

Here four--dimensional spacetime
corresponds to the coordinates $r, t_L, \theta$ and
$\phi$ while the two compactified dimensions
correspond to $t_R$ and $\zeta$.
The additional gauge field $A'_{\mu, \zeta}$ may be truncated
as in \btrunc, and $A'_{\mu, t_R}$ may be truncated in
an analogous way by setting
\eqn\bttrunc{
D_{t_R} = {\rm const,} \qquad
A'_{\mu, t_R} = \pm e^{2 D_{t_R}} A^{t_R}_{\mu}~.
}
The four--dimensional spacetime fields are then
(for simplicity we write
these only for region (i) of $SL(2 ,\IR)$ )
\eqn\rbfields{
\eqalign{
ds^2 &= {1\over 4} k_{SL} dr^2 - {1\over 4} k_{SL} \sinh^2 r d t_L^2 +
{1\over 4} k_{SU} d\theta^2 + {1\over 4} k_{SU} \sin^2 \theta d\phi^2 ~,\cr
A^{t_R}_{t_L} &= \gamma \cosh r,
\qquad A^{\zeta}_{\phi} =
m \cos \theta , \qquad
\phih =  {\rm const} \cr
e^{2D_{t_R}} &= {{k_{SL}}\over {4\gamma^2}},
\qquad e^{2D_{\zeta}} = {n \over {4m}}, \qquad
H_{\mu \nu \rho} =0 ~.\cr}
}
The central charge should again be set to 6, which
leads to $k_{SL}= k_{SU}+4$.

As described in \kallosh\ this solution possesses $N=2$ supersymmetry
when embedded in $N=4$, $d=4$ supergravity which in turn
is the low energy effective theory of the heterotic string
with the $E_8 \times E_8$ gauge fields truncated.
This corresponds to $(4,0)$ worldsheet supersymmetry. In terms
of the CFT, this is achieved by supersymmetrizing the WZW models
by adding free fermions. The $(4,0)$ supersymmetry algebra
is then constructed as in
\nref\sevrin{A. Sevrin, W. Troost, and
A. Van Proeyen, \PL {\bf 208B} (1988) 447.}%
\nref\callan{C.G. Callan, J.A. Harvey and A. Strominger,
\NP {\bf B359} (1991) 611.}%
\refs{\sevrin, \callan}.

\bigskip
{\bf Acknowledgements:}

D.~L. was supported in part by NSF
grant PHY-91-16964, and A. S.  was supported in part by DOE grant
91-ER40618.

\listrefs
\end